\documentclass[journal]{IEEEtran}

\pdfoutput=1

\usepackage{amsmath,bm,amssymb,amsfonts,amsthm}
\usepackage[linesnumbered, ruled, vlined]{algorithm2e}
\usepackage{graphicx}
\usepackage{xcolor} 
\usepackage{relsize}
\usepackage{textcomp}
\usepackage{float}
\usepackage[section]{placeins}

\usepackage{listings}
\usepackage{gensymb}

\usepackage{changepage}
\usepackage{hhline}
\usepackage{multirow}
\usepackage{nomencl}
\usepackage{mathtools}
\usepackage{commath}
\usepackage{booktabs}
\usepackage{subfiles}
\usepackage{tabularx}
\usepackage{lscape}
\usepackage{rotating}

\usepackage{verbatim}
\usepackage{comment}

\usepackage[normalem]{ulem}
\usepackage[utf8]{inputenc}
\usepackage[hyphens]{url}
\usepackage{longtable}
\usepackage{lineno}
    \modulolinenumbers[5]
\usepackage[caption=false]{subfig}
\usepackage{wrapfig}
\usepackage{enumitem}
    \setlist{nolistsep}
\usepackage[hidelinks]{hyperref}
\usepackage{makecell}
\usepackage{multicol}
\usepackage{colortbl}

\usepackage{scalerel}
\usepackage{tikz}
\usetikzlibrary{svg.path}
\definecolor{orcidlogocol}{HTML}{A6CE39}
\tikzset{
  orcidlogo/.pic={
    \fill[orcidlogocol] svg{M256,128c0,70.7-57.3,128-128,128C57.3,256,0,198.7,0,128C0,57.3,57.3,0,128,0C198.7,0,256,57.3,256,128z};
    \fill[white] svg{M86.3,186.2H70.9V79.1h15.4v48.4V186.2z}
                 svg{M108.9,79.1h41.6c39.6,0,57,28.3,57,53.6c0,27.5-21.5,53.6-56.8,53.6h-41.8V79.1z M124.3,172.4h24.5c34.9,0,42.9-26.5,42.9-39.7c0-21.5-13.7-39.7-43.7-39.7h-23.7V172.4z}
                 svg{M88.7,56.8c0,5.5-4.5,10.1-10.1,10.1c-5.6,0-10.1-4.6-10.1-10.1c0-5.6,4.5-10.1,10.1-10.1C84.2,46.7,88.7,51.3,88.7,56.8z};
  }
}
\newcommand\orcidicon[1]{\href{https://orcid.org/#1}{\mbox{\scalerel*{
\begin{tikzpicture}[yscale=-1,transform shape]
\pic{orcidlogo};
\end{tikzpicture}
}{|}}}}

\usepackage[noadjust]{cite}

\begin{document}

\title{\huge Measuring and Analyzing Effects of HEMP Simulation \\ on Synthetic Power Grids}

\author{
    Carson~L.~May $^{1}$\orcidicon{0000-0002-6580-1218},
    Arthur~K.~Barnes $^{2}$\orcidicon{0000-0001-9718-3197},
    Jose~E.~Tabarez $^{2}$\orcidicon{0000-0003-4800-6340}, \\
    Adam~Mate $^{2}$\orcidicon{0000-0002-5628-6509},
    Eric~M.~Nelson $^{3}$\orcidicon{0000-0003-1446-6453}, and
    Ross~Guttromson $^{4}$\orcidicon{0000-0000-0000-0000}
    \vspace{-0.1in}

\thanks{Manuscript submitted:~Nov.~30,~2022. Current version:~Jan.~30,~2023.
}

\thanks{$^{1}$ The author is with the Whiting School of Engineering, Johns Hopkins University, Baltimore, MD 21218 USA. Email: cmay16@jh.edu.}

\thanks{$^{2}$ The authors are with the Analytics, Intelligence, and Technology Division at Los Alamos National Laboratory, Los Alamos, NM 87545 USA. Email: abarnes@lanl.gov, jtabarez@lanl.gov, amate@lanl.gov.}

\thanks{$^{3}$ The author is with the Computational Physics Division at Los Alamos National Laboratory, Los Alamos, NM 87545 USA. Email: enelson@lanl.gov.}

\thanks{$^{4}$ The author is with the Electrical Power Systems Division at Sandia National Laboratory, Albuquerque, NM 87185 USA. Email: rguttro@sandia.gov}

\thanks{LA-UR-22-32309. Approved for public release; distribution is unlimited.}

\thanks{978-1-6654-9071-9/23/\$31.00 \copyright2023 IEEE}

}

\markboth{TPEC 2023 --- 2023 IEEE Texas Power and Energy Conference, February~2023}{}

\maketitle


\begin{abstract}
There is significant uncertainty about the potential effects of a high-altitude electromagnetic pulse (HEMP) detonation on the bulk electric system.
This study attempts to account for such uncertainty, in using Monte-Carlo methods to account for speculated range of effect of HEMP contingency. Through task parallelism and asynchronous processing techniques implemented throughout simulation, this study measure the effects of 700 large-scale HEMP simulations on a 7173 bus synthetic power grid. Analysis explores how contingency severity varies, depending on initial contingency parameters. Severity indices were captured throughout simulation to measure and quantify the cascading nature of an HEMP event. 
Further development of HEMP simulation modeling is explored as well, which could augment forecasts of potential contingency events as well.
\end{abstract}


\section{Introduction} \label{sec:introduction}
\indent


HEMP is low-probability, high consequence event which could cause significant disruption to the power grid through E1 and E3 coupling \cite{hutchins2016power}.
The E1 EMP component has high amplitude and frequency content but is of short duration. E1 EMP can couple into lines on the order of meters to tens of meters, resulting in primary effects such as permanent damage or temporary upset of a diverse set of components, such as generation plants, protective relaying on transmission and distribution systems, and customer loads \cite{ianoz1996modeling}.
The E3 EMP component has lower amplitude and frequency content but is of long duration, on the order of seconds to tens of seconds. E3 EMP couples into transmission lines on the order of tens to hundreds of km. By causing half-cycle saturation of large power transforms via inducing quasi-dc currents, E3 EMP can overwhelm the ability of generation to provide voltage support. This will in turn cause secondary effects, such as the loss of dynamic stability on the electrical grid and the potential for cascading events resulting from generation and line overloads \cite{horton_magnetohydrodynamic_2017, overbye2022towards}.
To predict impact, it is necessary to combine E1 and E3 coupling calculations with a device fragility model and dynamic simulation, and apply these in a Monte-Carlo framework to predict a statistical distribution of impacts \cite{timko1983monte}. 


Although the electromagnetic environments created from HEMPs differ from the environments created by geomagnetic disturbances (GMD), both in waveforms and time duration, the coupling equations for GMD can be applied to E3 environments \cite{overbye_ground, mate21-pmsgmd}.
The induced currents from both E3 and GMD environments are calculated based on Faraday’s law, which describes the coupling of the electrical fields onto transmission lines. These currents can lead to overloading of lines and transformer saturation resulting in the operation of protection devices within the model opening up breakers. This can be modeled numerically.
Starting with version 20, PowerWorld Simulator added the ability to import HEMP environments through the GIC toolbox to incorporate into transient analysis \cite{powerworld_e3}. The induced currents are calculated at each time step during the analysis, but because of the size of the event and complexity of electrical models non-convergence can become an issue.
One source of possible non-convergence is due to the discontinuities in the dynamic equations created by breaker operations \cite{overbye2022towards}. This limits the ability to fully investigate the effects that E3 environments have on the power system. Additionally, on account of the lack of experimental data and the large number of effected devices, interactions and timescales, it is hard to predict effects. With ``conventional'' effects of nuclear weapons, there is significant testing data and open literature to predict impacts. This has been used to produce simple ``slide-rule'' type calculations for severity indices such as static or dynamic overpressure \cite{glasstone1977effects}.
This motivates the work presented here, where we look for simple univariate relationships to predict the impact of HEMP based on simulation.

Performing end-end coupling and dynamic simulation of EMP, as described earlier, is challenging because of data and computational challenges.
However, network science \cite{cicilio21-resilience, gorham2019identifying, hines2017cascading} and forecasting approaches \cite{tang2018framework, xie2020review} have gained increasing popularity for assessing power systems resilience based on characteristics of the network.
This could include graph properties such as assortativity or electrical properties, such as the amount of generation surplus.
At the distribution level, statistical machine learning has been successfully applied to predict hurricane impacts \cite{han_estimating_2009, mcroberts2018improving} without the need for a solveable electrical model of the distribution network.
This work looks for relationships with network characteristics and severity indices, in order to gain insight about what impacts vulnerability of networks to HEMP and to guide insight of future work for developing forecasting capabilities.


This study aims to measure, analyze, and visualize the various effects of high-altitude electromagnetic pulse (HEMP) detonation on the U.S. electrical grid.
The intent is to provide insight as to how a detonation could realistically impact a power system, given parameters such as detonation size and location. This could lead to the prediction of HEMP effects on a grid, by identifying any correlated measurements or states of the electrical network that contribute to the cascading impact from the HEMP detonation.


This paper builds on the work of \cite{gorham2019identifying} to consider the impact of network properties on severity indices associated with a cascading event.
Simulation began with line failure induced throughout E1 waveform emulation, as ac power flow processed the effects on a large scale synthetic power grid. This process serves as a proxy for transient stability.
A parallelized implementation enabled over 700 simulations to be performed on a 7173-bus synthetic grid over the course of three days. To measure results, severity indices such as line outage distribution factor (LODF) and generation surplus were analyzed. Here, both the distribution and spatial variation of the the severity indices are considered along with correlation of the severity indices and other quantities.


The remainder of this paper is structured as follows:
Section~\ref{sec:methodology} introduces the developed Monte Carlo HEMP simulation model, describes the synthetic ERCOT power system model used for the case studies, and establishes measurements and parameters used for performance analysis.
Section~\ref{sec:results} presents the results of the HEMP simulation through measures of the severity indices and contingency effects.
Lastly, Section~\ref{sec:conclusions} summarizes findings and describes potential future developments that could improve contingency modeling and develop general predictive capabilities of contingency effects.

\section{Methodology} \label{sec:methodology}
\indent

\subsection{HEMP simulation model}

A randomly generated contingency scenario was developed that notionally represents the damage to an electrical grid resulting from a HEMP detonation.
The contingency defines a random set of transmission line failures, where the probability of a line failure is most likely at ``ground zero'' of the detonation and decreasing out to a specified damage radius.
This model is a prototype; a more realistic damage model is currently under development. This simulation was generated from a customization of the HEMP Transmission Consequence Model (HTCM), developed under eponymous HTCM project. 

The HTCM based HEMP simulation model, in this instance, simulates the effects of the early-time E1 waveform on a synthetic power system model located in the State of Texas. The times of the line outages within the damage contingency radius are determined by a Gamma probability distribution.
As stated above, the likelihood of a line failure is spatially dependent on its location to ``ground zero''; it ranges from a maximum failure likelihood at the contingency center to zero probability at the contingency radius.

Probability of line failure, within the affected radius, was designated to occur when

\begin{figure}[!htbp]
\centering
\includegraphics[width=0.45\textwidth]{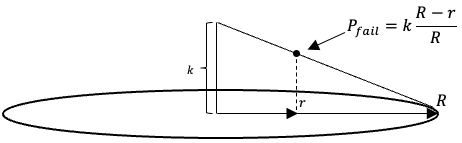}
\caption{Developed probabilistic effect HEMP model.}
\label{fig:probabilistic-model}
\end{figure}

\small
\begin{equation}
U \le k\left( \frac{R-r}{R} \right)
\label{eq:failure-prob}
\end{equation}
\normalsize \noindent where
$U$ is a random variable defined by a uniform pdf over $[0, 1]$;
$k$ is the slope of failure probability;
$R$ is the probabilistic effect radius; and
$r$ is the distance of the transmission line from the point of contingency.
Fig.~\ref{fig:probabilistic-model} visualizes these parameters of the calculation.

Effects of contingency-induced transmission line failures may initiate cascading events in one or more places across the electrical grid. This is critical because existing work has demonstrated that component failures during a cascading event may be nonlocal \cite{hines2017cascading}.
For the simulations performed in this study, an initial probabilistic effected radius of 100 km was placed on a synthetic 6717-bus model of the Texas electrical network \cite{li2020building}. The simulations performed a HEMP detonation at the centroids of each of the 254 Texas counties, with 3 different line failure probability simulations performed at each centroid; in total, 762 contingency simulations were performed.
Simulation data was collected for one line failure at a time-step before non-convergence of the power flow; that is, if one more line were to fail during steady-state power flow simulation, the power flow would not converge.

The number of line failures until non-convergence was assessed using PowerWorld Simulator 22's AC power flow solver, with generator active and reactive power limits enabled.
Transmission lines were iteratively disabled based on the line failure times defined above. At each iteration, 5 lines were disabled and the PowerWorld AC power flow solver was run to determine if the system was converged to a feasible solution.
After each successful solve, the system state was saved so that the system state just prior to non-convergence could by studied.
To accelerate the time it takes to perform all of the simulations, separate PowerWorld instances were run in parallel for each scenario associated with a particular contingency origin point.

\subsection{Nature of Synthetic Grid}
\indent

The developed HEMP simulation model was deployed on the Texas7k synthetic power system model.
The model was designed to mimic the geographic footprint of the Electric Reliability Council of Texas (ERCOT) supervised electrical grid. It has similar characteristics to the actual grid but without the potential communication of confidential information \cite{Birchfield2016, Birchfield2018}. 



\subsection{Predictors: Line Outage Distribution Factor and Generation Surplus}
\indent

For this study, the \textit{line outage distribution factor} (LODF) and \textit{generation surplus} (GS) features were used to measure the effect of contingency simulations.

LODF is described by \eqref{eq:lodf}; it specifies a percentage of increase in power flowing in transmission line \emph{n} when line \emph{i} experiences an outage \cite{wood2013power}. For this study, the lines were counted via \eqref{eq:lodf_count} to identify cases where many lines experienced an active power increase over a specified percentage threshold. \eqref{eq:lodf_count} provides the logic-check used to generate the count for each line, were a single line to open in a functioning power system is:

\small
\begin{equation}
LODF_{n} = \frac{ \Delta P_{n}}{\Delta P_{i}}
\label{eq:lodf}
\end{equation}
\begin{equation}
SI = Count \left(\frac{ \Delta P_{n}}{\Delta P_{i}} \ge 3\%  \; \forall \: i  \in N_{line},\: i \ne n \right).
\label{eq:lodf_count}
\end{equation}
\normalsize \noindent This feature can be applied to a geographical assortment of lines to determine areas in which line openings have particularly strong, and potentially cascading, effects on power flow.

GS is described by \eqref{eq:gensurpl}; it is calculated on a per generator basis as:

\small
\begin{equation}
GS = GenMaxPower \left(\frac{GenMWPercent - 100}{100}\right)
\label{eq:gensurpl}
\end{equation}

\normalsize
\section{Results} \label{sec:results}
\indent
\subsection{Bus and Line Failures Throughout Contingency}

Two additional features were introduced to analyze the effects of the developed HEMP simulation model: \textit{bus failure} (BF) and \textit{line failure} (LF).
BF specifies the total number of bus outages at the timestep just prior to non-convergence was measured; it is a measure of the number of bus outages at a particular timestep that are selected to fail as a result of \eqref{eq:failure-prob} and the cascading failure that follows.
LF specifies the total number of transmission lines opened throughout a contingency event; it is a measure of line outages induced by the event, and when visualized, can communicate the geographic implications of the contingency.
Results of both measures vary quite dramatically by contingency.

\begin{figure}[!htbp]
\centering
\includegraphics[width=0.475\textwidth]{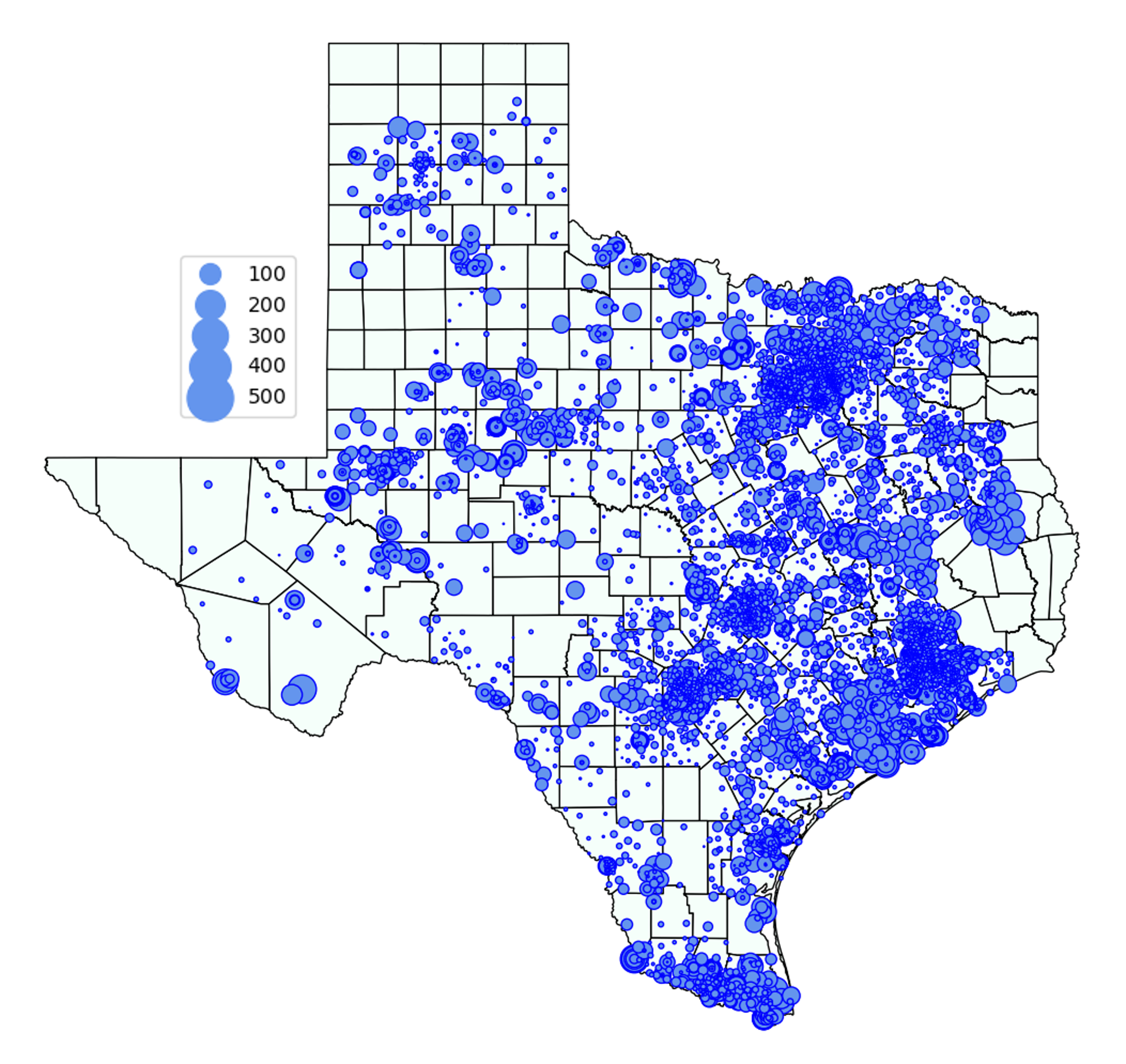}
\caption{Cumulative BFs across 762 contingency scenarios.}
\label{fig:bus-opening-count}
\end{figure}

Fig.~\ref{fig:bus-opening-count} provides a geographic representation of cumulative BF measures across all 762 contingency scenarios.
It must be noted that the neither the geographic nor the electrical proximity of buses appear to correlate with their number of cumulative failures.

\begin{figure}[!htbp]
\centering
\includegraphics[trim=0 0 0 1.39cm,
clip,width=0.45\textwidth]{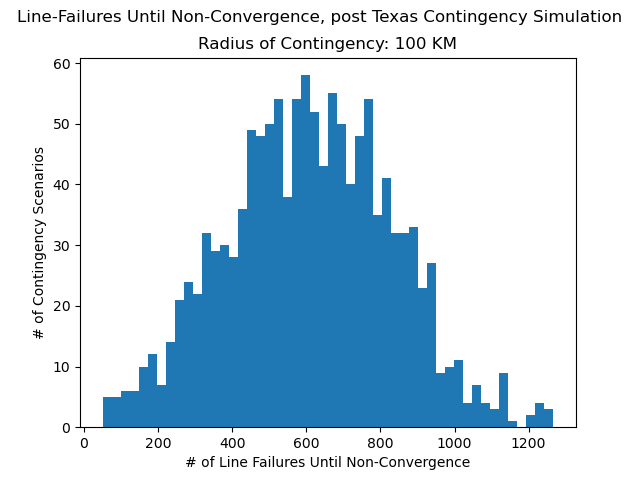}
\caption{Number of LFs until non-convergence for 762 contingency scenarios.}
\label{fig:branch-failures-to-nonconvergence}
\end{figure}

Fig.~\ref{fig:branch-failures-to-nonconvergence} illustrates the distribution of LF measures until non-convergence, across all 762 scenarios.
The total number of failures follows a Gaussian distribution with a mean of approximately 650 transmission lines. This is approximately 10\% of the total number of lines in the system.

\begin{figure}[!htbp]
\centering
\includegraphics[width=0.50\textwidth]{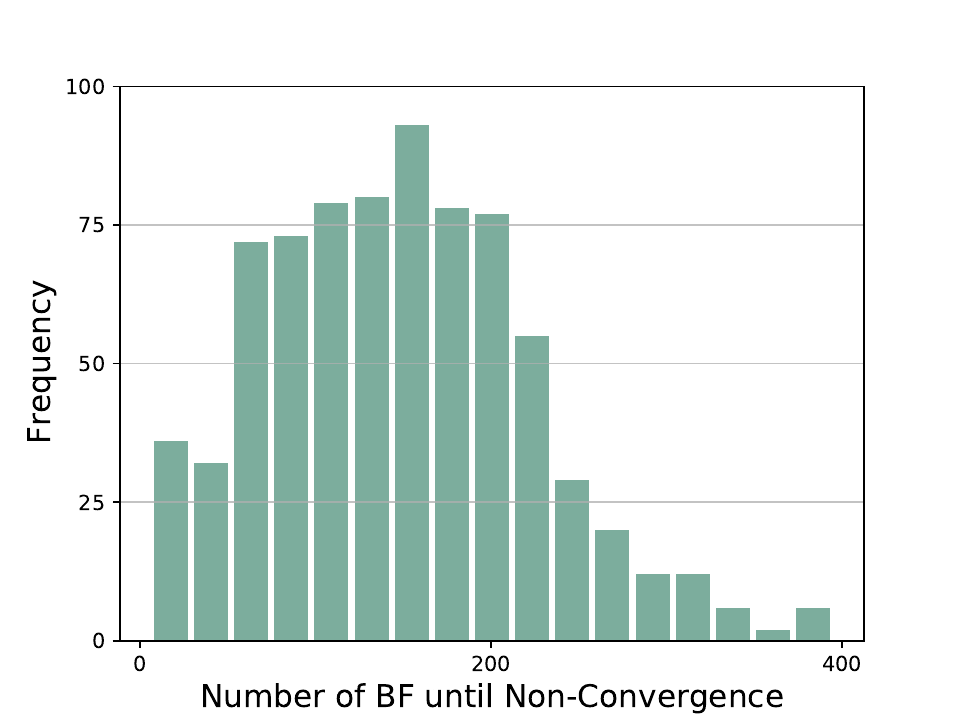}
\caption{Total number of BFs throughout contingency, by scenario.}
\label{fig:buses-opened-per-scenario}
\end{figure}

Fig.~\ref{fig:buses-opened-per-scenario} provides a summation of BF measures by contingency scenario, caused by the initiation of bus protection schemes throughout a contingency.

\begin{figure}[!htbp]
\includegraphics[trim=0 0 0 0, clip,width=0.475\textwidth]{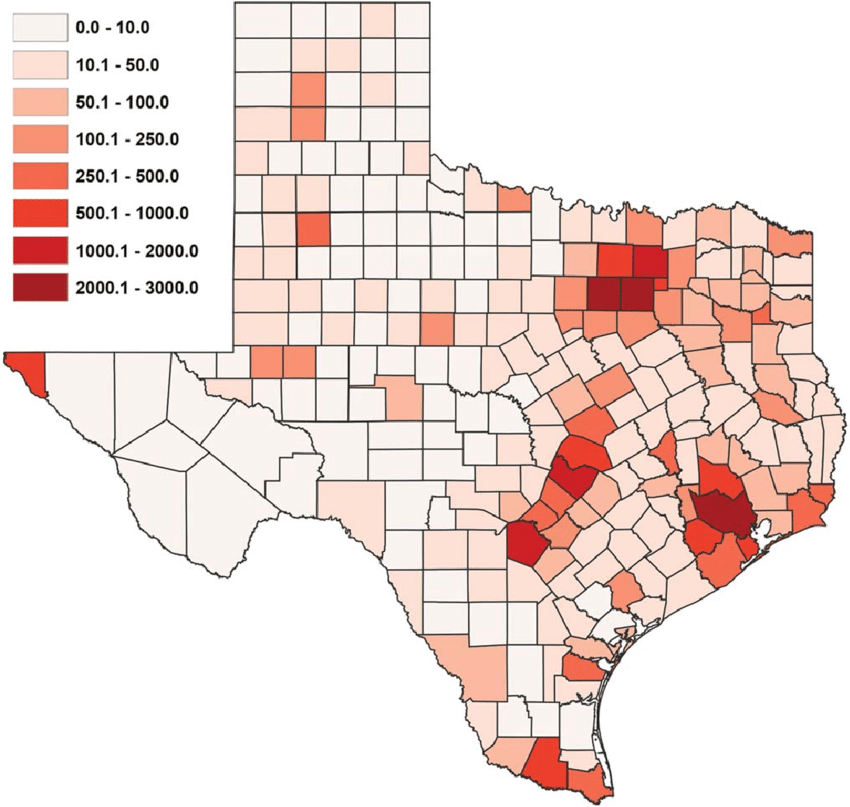}
\caption{Population density of Texas counties (people/square mile).}
\label{fig:texas-density}
\end{figure}

The assumption that population density correlates with total line failure caused by contingency was investigated: one 50~km contingency scenario was performed for each county centroid.
Figs.~\ref{fig:line-failures-to-nonconvergence-vs-pop-density} and \ref{fig:texas-density} show the LF measures in each county, and the approximate population density of counties: 

\begin{figure}[!htbp]
\includegraphics[trim=0 0 0 0, clip,width=0.475\textwidth]{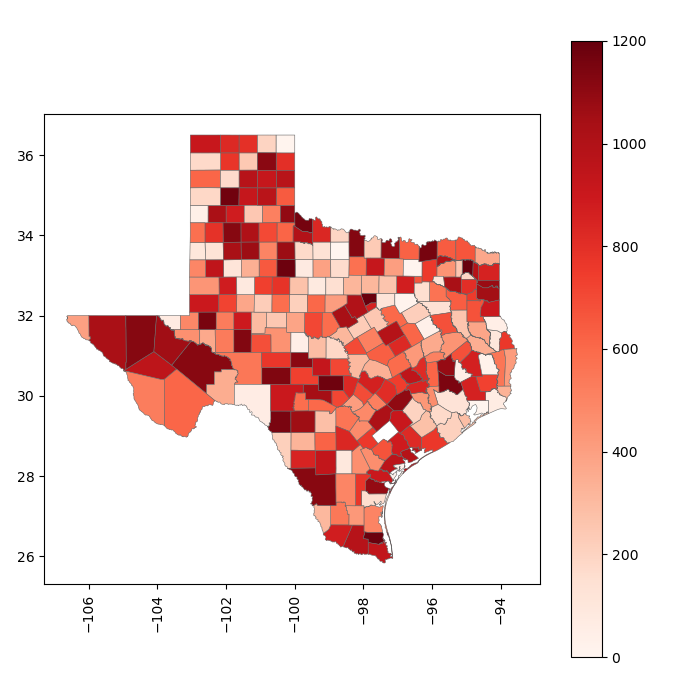}
\caption{Number of LFs  until non-convergence, per contingency initialized in county.}
\label{fig:line-failures-to-nonconvergence-vs-pop-density}
\end{figure}

When plotted, data confirms little to no correlation between the two measures, as shown in Fig.~\ref{fig:corr-pop}:

\begin{figure}[!htbp]
\includegraphics[trim=0 0 0 0, clip,width=0.40\textwidth]{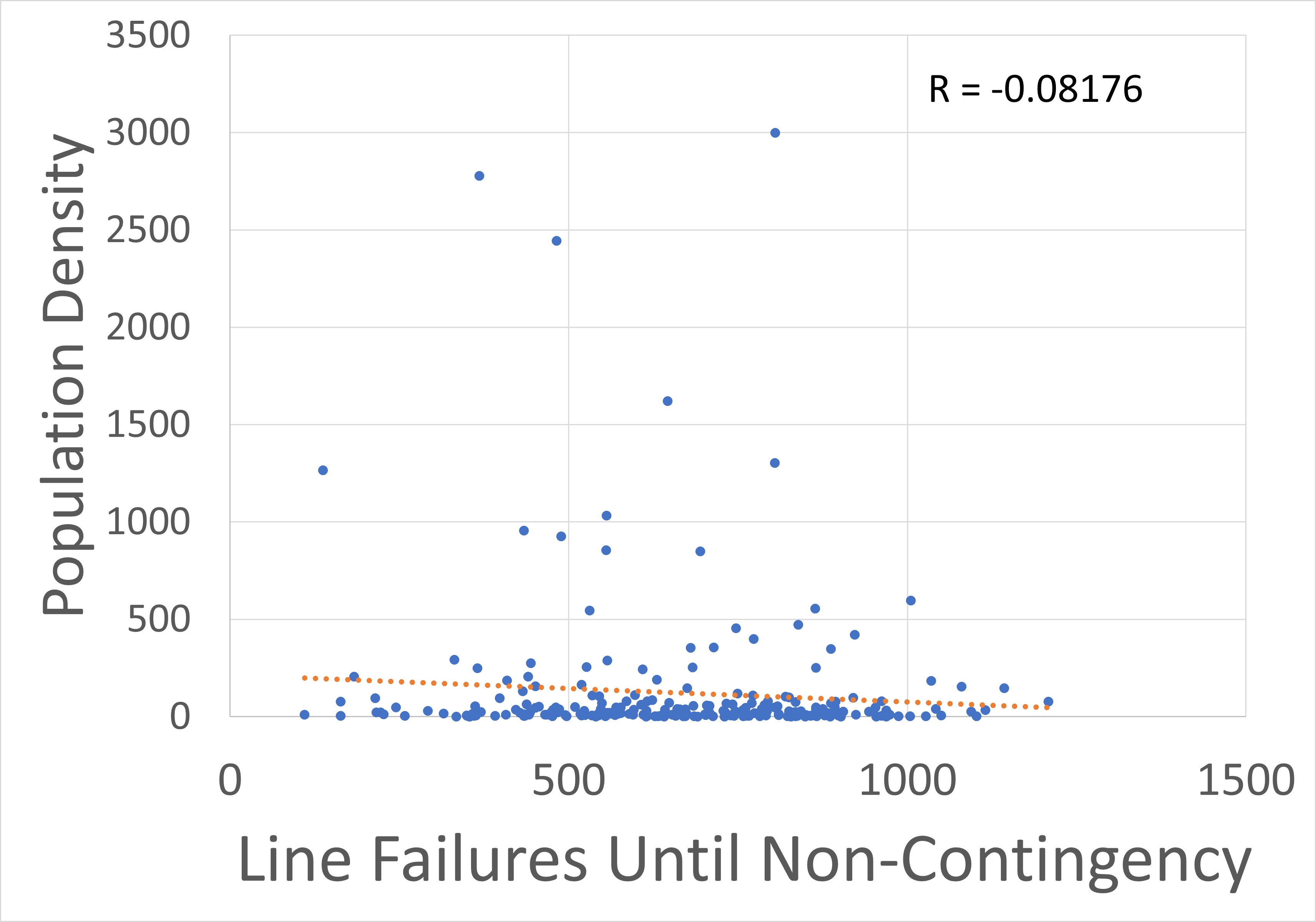}
\caption{Number of LFs until non-convergence (X) versus population density of county (Y).}
\label{fig:corr-pop}
\end{figure}


\subsection{Line Outage Distribution Factor}
\indent

\begin{figure}[!htbp]
\centering
\includegraphics[trim=32cm 0 0 0, clip,width=0.65\textwidth]{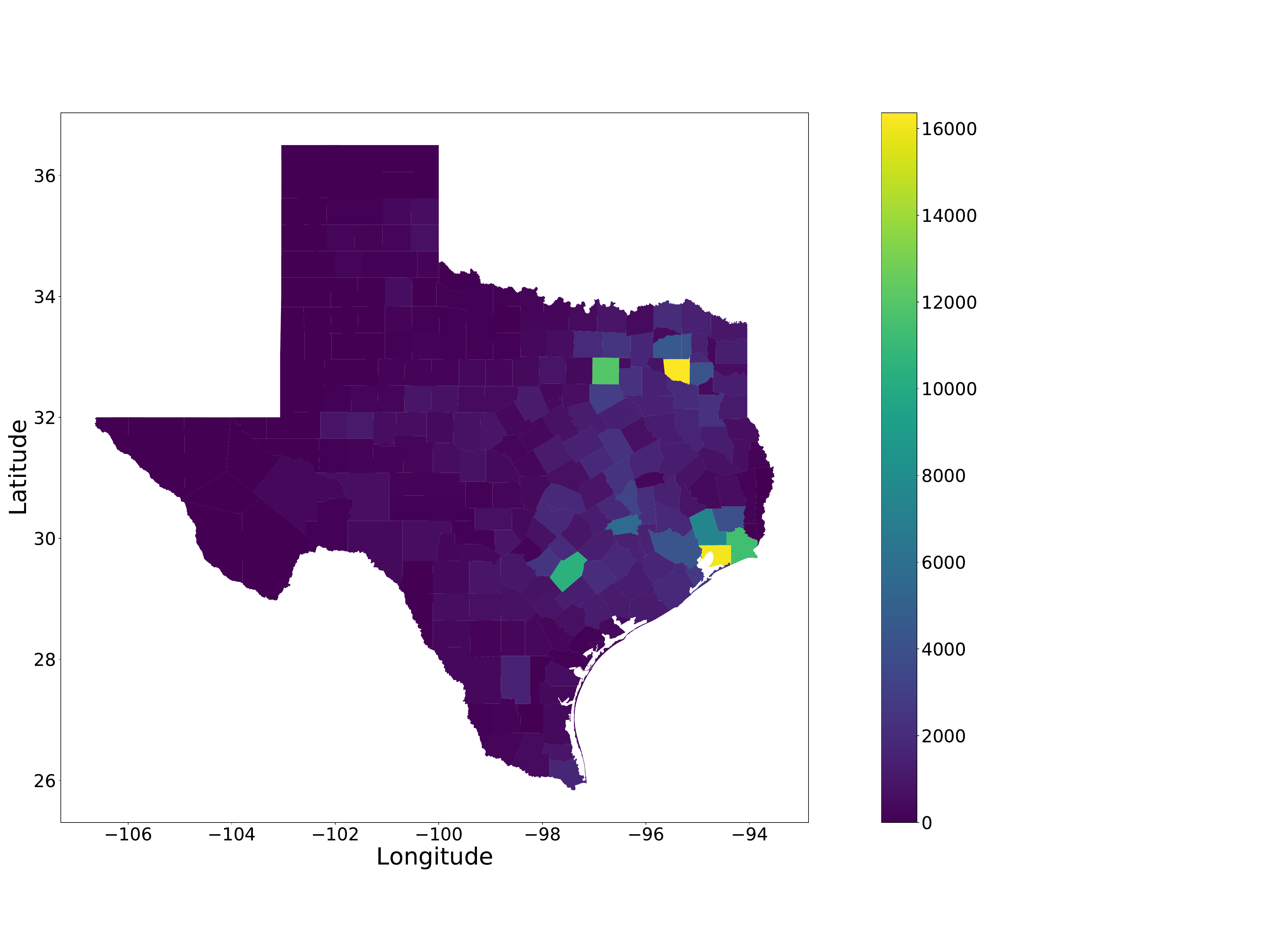}
\caption{Total number of lines experiencing greater than 3\% power flow increase due to outage of other line in county.}
\label{fig:lodf-pre-contingency}
\end{figure}

LODF measures how a line opening affects the power flow on other lines within the electrical network. Counting the number of lines that experience a specified increase in percentage of active power, functions as a severity index: areas with particularly high values could be particularly vulnerable to HEMP effects.
In Fig.~\ref{fig:lodf-pre-contingency}, the severity indices are summed on a per county basis; this is done pre-contingency, to indicate potential vulnerabilities before outages occur.

This feature can be normalized by dividing the county total by the number of lines within the county; this essentially indicates, on average, the number of lines that experience greater than 3\% LODF due to a line outage in the county.
Fig.~\ref{fig:lodf-post-contingency-normalized} visualizes this normalization:

\begin{figure}[!htbp]
\centering
\includegraphics[trim=31cm 0 0 0, clip,width=0.65\textwidth]{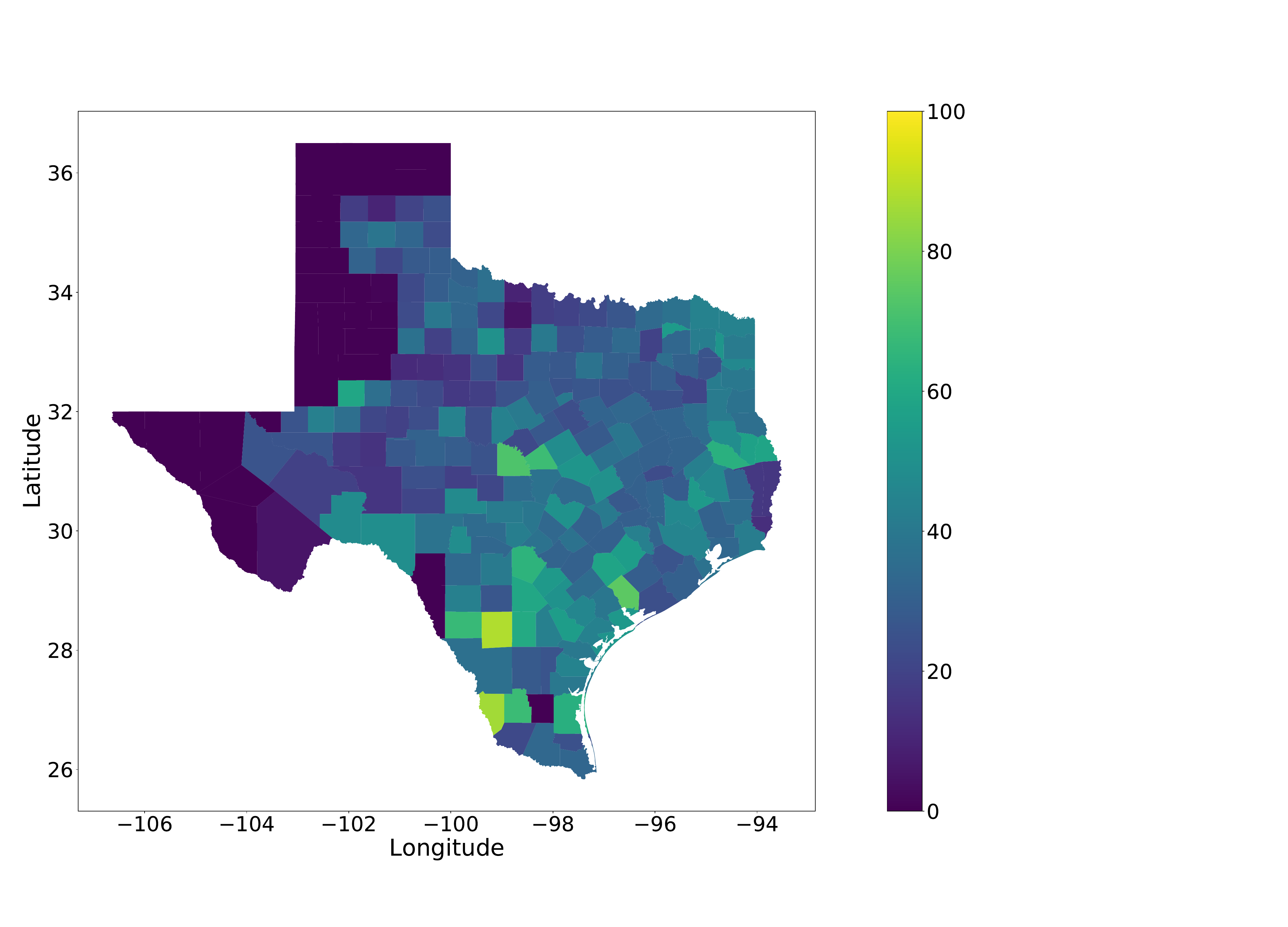}
\caption{Average number of lines experiencing greater than 3\% LODF, per outage, in county.}
\label{fig:lodf-post-contingency-normalized}
\end{figure}


\subsection{Generation Surplus}
\indent

\begin{figure}[H]
\centering
\includegraphics[width=0.48\textwidth]{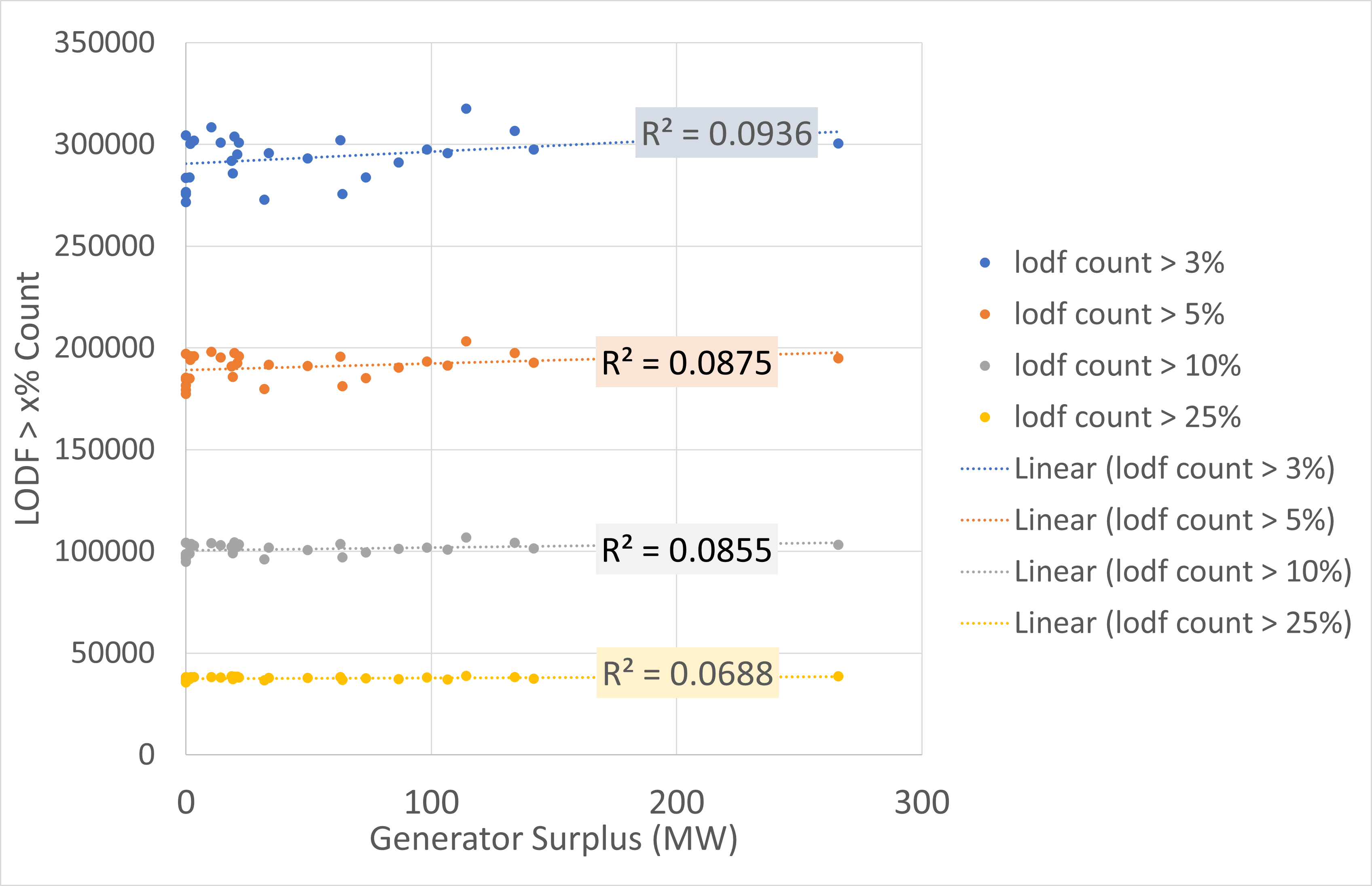}
\caption{GS [MW] versus LODF SI totals, per contingency.}
\label{fig:gen_vs_lodf}
\end{figure}

GS was measured one line failure before non-convergence and compared to calculated LODF features -- LODF counts were found through the application of \ref{eq:lodf_count} at different LODF thresholds -- across the contingencies. Altogether thirty simulations from thirty randomly selected counties, were performed.
Fig.~\ref{fig:gen_vs_lodf} indicates that higher variation in GS can be explained by lower threshold LODF counts; though it must be noted that the correlation is relatively low. This is desirable as it means that the two quantities are independent.

\begin{figure}[H]
\centering
\includegraphics[width=0.48\textwidth]{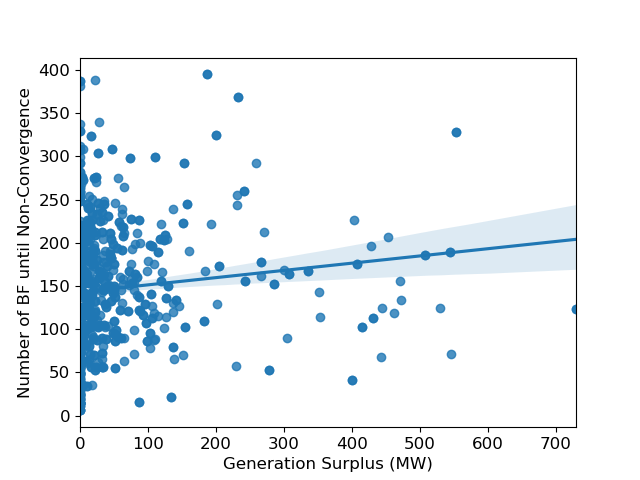}
\caption{Correlation between GS [MW] and LODF totals, per contingency.}
\label{fig:gen-surplus-vs-dead-bus-count}
\end{figure}

Fig.~\ref{fig:gen-surplus-vs-dead-bus-count} suggests a slight correlation between GS and the total number of BF in the contingency.

\subsection{Validation}
\indent

In order to ensure the accurate modeling of the simulated contingencies and that data was not corrupted by unrealistic non-convergence results, key characteristics of the contingency events were examined.

\begin{figure}[!htbp]
\centering
\includegraphics[width=0.48\textwidth]{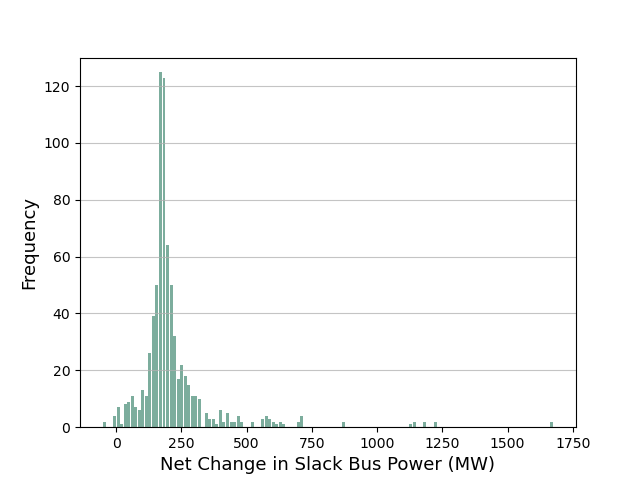}
\caption{Net change in active power drawn by the slack bus, across all contingency scenarios.}
\label{fig:slack-bus-dp}
\end{figure}

\begin{figure}[!htbp]
\centering
\includegraphics[trim=0 0 0 1cm, clip, width=0.48\textwidth]{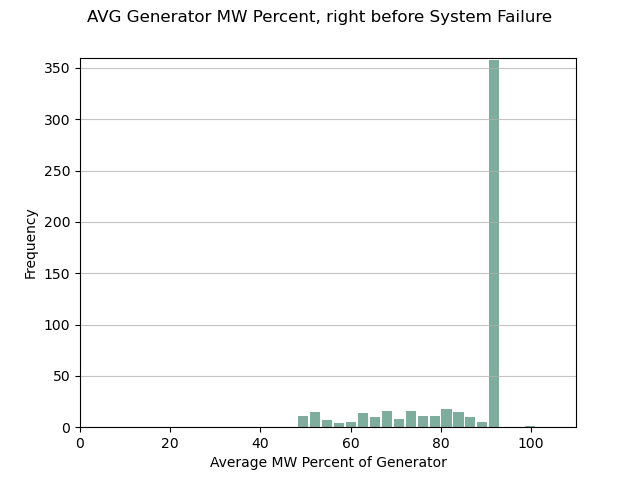}
\caption{Distribution of average generation loading right before non-convergence, across all contingency scenarios.}
\label{fig:gen-loading-distn-pre-conconvergence}
\end{figure}

Fig.~\ref{fig:slack-bus-dp} displays the distribution of active power drawn by the slack bus, while Fig.~\ref{fig:gen-loading-distn-pre-conconvergence} displays the distribution of generation loading right before non-convergence, across all 762 scenarios.
These figures illustrate that power flows are in a reasonable state prior to reaching non-convergence power flow as the slack bus active power is not exceptionally large and the majority of generators remain at peak loading.

\normalsize
\section{Conclusions} \label{sec:conclusions}
\indent

Despite the complexities of simulating a HEMP event on an electrical grid approximating ERCOT, no features indicate issues with the steady-state power flow performed throughout the contingency event.
Results indicate how critical geography and grid structure are to contingency severity, as line failures and bus failures vary quite dramatically by scenario, as seen in Fig.~\ref{fig:branch-failures-to-nonconvergence} and Fig.~\ref{fig:buses-opened-per-scenario}. However, results do not indicate that population density near contingency initialization is related to contingency severity, as seen in Fig.~\ref{fig:corr-pop}. This finding suggests that the effect of cascading failure is often wide-spread, and that contingency severity is not simply a reflection of power consumption within initial contingency radius.
Though results are promising, there is a lot more to be done in this space.

The damage model employed for simulation is a surrogate, and a higher fidelity damage model is in development under the HTCM project. The developed probabilistic effect model, used for HEMP simulation, could be upgraded to take additional parameters such as late-time E3 EMP induced currents effects, line angling, and geomagnetic effects into account. Different EMP parameters could be considered as well, such as detonation height and initial contingency effect radius. 
Additional HEMP severity indices could also be considered to better measure and predict HEMP effects.
Lastly, implementation of ML approaches shows promise as well; the next subsection details this proposition in more detail.

Additional features -- including severity indices, topological aspects of electrical grids, and other power systems calculations -- have been found to be useful to predict and analyze the effects of HEMP events.
For example, various graph theory features have shown to have moderate to high correlation with power shed due to a contingency \cite{gorham2019identifying}.

Analysis of simulation results obtained throughout this study has found additional features that show high correlation with HEMP effects. 

These new features, alongside with established features, may be used to train ML algorithms to anticipate the effects of potential HEMP events on the U.S. electrical grid. This would enable predictions of HEMP effects to be quickly acquired, without running lengthy simulations.
The general predictive ability of ML approached would also enable reasonably accurate predictions for contingencies other than HEMP, such as hurricanes, ice/wind storms, targeted intelligent adversary attacks, or widespread cyberattacks.


\bibliographystyle{unsrt}
\bibliography{references}

\end{document}